\documentclass[12pt,a4paper]{article}
\pdfoutput=1
\usepackage{epsfig,amsmath,amssymb}
\usepackage{natbib}
\usepackage{bm, url}
\usepackage{psfrag}
\usepackage{wrapfig}
\usepackage{booktabs}
\usepackage{geometry}
\usepackage{pdflscape}
\usepackage[width=0.9\textwidth]{caption}

\def\e{\hbox{E}}

\def\var{\hbox{Var}}

\begin{document}

\title{Insights and inference for the proportion below the relative poverty line}
\author{Dilanka S. Dedduwakumara \textsuperscript{a}, Luke A. Prendergast \textsuperscript{a} and Robert G. Staudte \textsuperscript{a} \\ \\ \textsuperscript{a}Department of Mathematics and Statistics\\
La Trobe University, Melbourne, Australia}

\maketitle

\abstract{We examine a commonly used relative poverty measure called the headcount ratio ($H_p$), defined to be the proportion
of incomes falling below the relative poverty line, which is defined to be a fraction $p$ of the median income.  We do this by considering this
concept for theoretical income populations, and its potential for determining
actual changes following transfer of incomes from the wealthy to those whose
incomes fall below the relative poverty line. In the process we derive and evaluate
the performance of large sample confidence intervals for $H_p$. Finally, we illustrate
the estimators on real income data sets.}

\vspace{0.5cm}


 \section{Introduction}

Simple-to-interpret and scale-free measures of poverty are commonly used to assess economic health of a country, either in comparison to other countries, or as a measure of change within a country compared to historical measures.  One commonly used measure, often referred to as the `headcount ratio', is the proportion of individuals whose income is less than a fraction, $p$, of the median income.  Usually, $p$ is chosen to be between 40\% and 60\% \citep[e.g. see][who use these rates with $p$ equal to 40\%, 50\% and 60\% as measures of poverty in Germany and the United States]{burkhauser1996relative} and for many countries a specific choice of $p$ is used to define the official poverty line.  The Organisation for Economic Co-operation and Development (OECD) specifically defines the poverty rate using $p=0.5$ (\url{https://data.oecd.org/inequality/poverty-rate.htm}) and this has been adopted by many governments.  For example, in Hong Kong it is $p=0.5$ and recently this definition of poverty has been used to assess factors associated with poverty \citep{Peng2019}.  In the European Union the choice of $p=0.6$ is referred to as the `at-risk-of-poverty' rate and examples of this for many countries can be found in, for example, Figure 3.1 of \cite{Bradshaw2019}.

The purpose of this paper is two-fold.  Firstly, we provide some insights into this povertymeasure with respect to some probability distributions often used to model income.  Secondly, we provide simple confidence intervals that may be used as estimators when a sample of incomes is available.  We begin, in Section \ref{sec:definitions} with some formal definitions which are essential for clarity of analyses and inferences to follow before providing some properties and examples.  In Section \ref{sec:inference} we discuss inference, including estimators for both complete data sets and data summarised in grouped format.  Simulation studies to assess performance of estimators are also included.  Applications are in Section \ref{sec:Applications} before we conclude in Section \ref{sec:discussion}.

\section{Definitions, properties and some insights}\label{sec:definitions}

Let $F_\sigma (x)=F_1(x/\sigma )$, for all $x>0$ and some unknown continuous $F_1\equiv F$ and unknown scale parameter $\sigma >0 .$
If $X$ is a randomly chosen  income from this distribution, we write $X\sim F_\sigma ,$ and drop the $\sigma $ subscript when
there is no need to emphasize its presence. We further assume that the density $f(x)=F'(x)$ exists and is positive for all $x>0.$ Define for $0<u<1$ the  quantile function $Q(u)=\inf \{x: \  F(x)\geq u\}.$  Sometimes we abbreviate $Q(u)$ to $x_u$.
It can be shown that  $F(x_u)=u$ since $F$ is continuous. Let $M=x_{0.5}$ denote the unique median of $F$.

The {\em relative poverty line} for a given year is often defined to be $L_p=p\times M,$ for some fixed $p\in (0, 1)$ where $F$ is the income distribution during that year. The {\em poverty index} is then defined by $H_p=F(L_p).$

\subsection{Properties of $H_p$}

Below we list several properties of $H_p$ which has led many to use the measure as a simple-to-interpret measure of relative poverty.
\begin{description}
  \item[P1.]   $H_p$ has the interpretation of being the proportion of the population of incomes that are less than
  the relative poverty line $L_p$.  The larger $H_p$ is, the greater the relative poverty in the population.
  \item[P2.]  $0 \leq H_p $, where $H_p=F(L_p)=0$ when $L_p$ is less than the smallest income in the population; that is, when the support
   of $F$ lies to the right of $L_p=p\times M$. This last case can be interpreted as \lq zero relative poverty\rq.
  \item[P3.]  $H_p \leq 1/2$, where $H_p=F(L_p)=1/2$ only if there were no incomes between $L_p=p\times M$ and $M$. This last case would be
  interpreted as \lq maximum relative poverty\rq . \
  \item[P4.] $H_p$ is scale invariant. For if $X\sim F$ and $Y=cX$, where $c>0$ then $H_{p,Y}=H_{p,X}$. To see this, note that
  $m_Y=c\,m_X$, so $L_{p,Y}=c\,L_{p,X}$; and further $F_Y(y)=F_X(y/c)$ for all $y$.  Hence $H_{p,Y}=F_{p,Y}(L_{p,Y})=F_X(c\,L_{p,X}/c)=F_X(L_{p,X})=H_{p,X}.$
\end{description}
To summarise the above properties, $H_p\in [0,1/2]$ is a scale-free measure of relative poverty where $H_p=0$ and $H_p=1/2$ are interpreted as \lq zero relative poverty\rq and \lq maximum relative poverty\rq respectively.  While this gives the impression that $H_p$ is a useful measure of relative poverty, it is important to note that $H_p$ does not depend at all on incomes above the median.  Consequently, any event that results in an influx of wealth to the richer half of the population, while the poorer half remain steady, including the median, will not translate to increases in relative poverty as measured by $H_p$.


\subsection{Examples for several distributions}

For simplicity with these examples we assume that $p=0.5$ and use $H=H_{0.5}$ and $L=L_{0.5}$.  Results can similarly be derived for other choices of $p$.

\subsubsection{Uniform} While it is not often used to model income data, the uniform distribution does provide some interesting insights into the behavior of $H$.  For $X\sim \text{Unif}(a, b)$ with $a < b$, we have
$$H = \begin{cases} 
0 & (a+b)/4 < a \\
\frac{b-3a}{4(b-a)} & a \leq (a+b)/4 \leq b
\end{cases} $$
A special case is for when $a=0$ representing `zero income' and for which $H=1/4$ which is simple to verify intuitively by definition of $L$ being half the median.  For $a>0$ and $b=c\times a$ for some $c>0$, we see that $\lim_{c\rightarrow \infty}H=1/4$ which is the maximum value of $H$ for the uniform distribution.  Again, this intuitively makes sense due to the concept of uniformity and the proportion between $M/2$ and $M$ cannot be less than 1/4.  The uniform distribution also provides a convenient means to highlight the effect on $H$ if a proportion of population were shifted to higher income levels.

\begin{figure}[ht]
    \centering
    \includegraphics[scale=0.6]{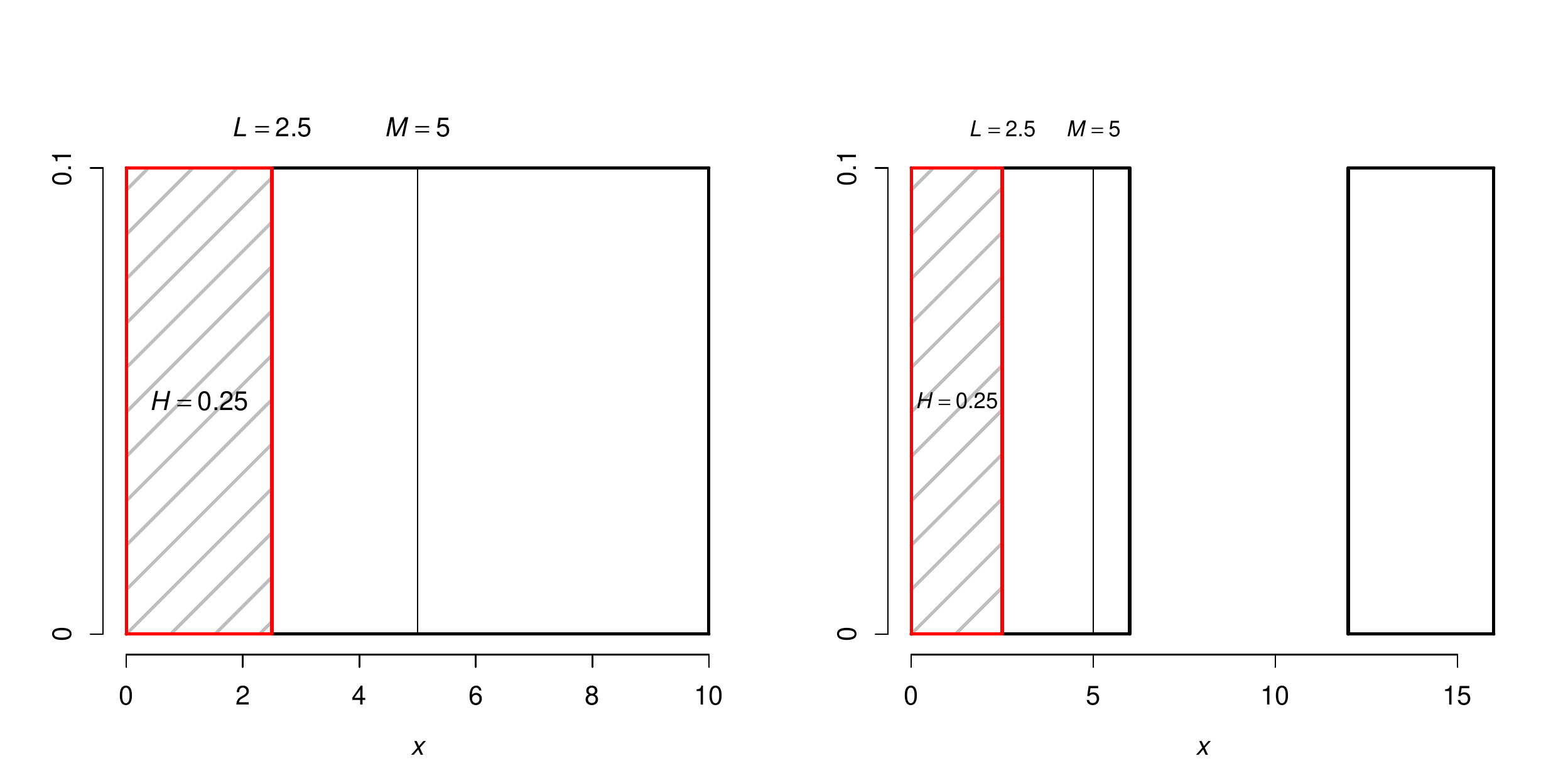}
    \caption{Example density functions for the Uniform$(0, 10)$ distribution (left) and the resulting density when all $x$ in the top 40\% of the density have the value 6 added.  Values of the median $(M)$, $M/2$ and $H$ are annotated on the figures.}
    \label{fig:unif}
\end{figure}

In Figure \ref{fig:unif} we plot two probability densities.  The left plot is the density for the Uniform$(0,10)$ distribution which, as noted above, has $H=0.25$ (see the shaded area).  In the right-side plot we change the density by shifting the highest 40\% of the probability mass 6 units to the right.  Despite this increase in `wealth' for the top 40\%, $H$ remains the same at 0.25 indicating no change in relative poverty.  This example can be repeated for any other distribution of wealth where the top 50\% of incomes have no influence on $H$.

\subsubsection{Lognormal} For the lognormal distribution $X\sim F(x)= \Phi (\ln (x))$ for $x>0$, where $\Phi $ is the standard normal distribution. The median is the solution of $\Phi (\ln (M))=0.5$, or $M=1$.  Hence $H=F(0.5)=\Phi (\ln (0.5))=\Phi (-\ln (2))= 0.244108.$

\subsubsection{Pareto}For the Type II Pareto($a$) distribution, $X\sim F(x)=1-(1+x)^{-a} $ for $x>0$ and shape parameter $a>0.$
      The median is the solution of $(1+M)^{-a} =0.5$, or $M=2^{1/a}-1$. Hence
      \[H(a)=F(M/2)=1-\{1+2^{(1-a)/a}-1/2\}^{-a}= 1-2^{a}\{1+2^{1/a}\}^{-a}~.\]
     By direct computation,  $H(1)=1/3$, $H(2)=0.3137$ and $H(6)=0.29993.$  A plot of $H(a)$ versus $a$ reveals that $H$ is monotone decreasing from $H(0+)=0.5$ to $\lim _{a\to \infty }H(a)=1-\sqrt {2}\,/2= 0.29289\approx 0.3.
   $

\subsubsection{Weibull} For the  Weibull$(b)$ distribution,  $X\sim F(x)=1- e^{-x^{b}}, $ for $x>0$ and some shape parameter $b >0.$ The median is the solution of $e^{-M^b } =0.5$, or $M=\{\ln (2)\}^{1/b }$. Hence
      \[H(b )=F(M/2)=1-\exp \{-(\{\ln (2)\}^{1/b }/2)^{b}\}  =1-2^{-2^{-b }}~.\]
           Some examples are $H(1)=0.29289$, $H(2)= 0.1591$ and $H(6)=0.0108.$  $H(b)$ is monotone decreasing from $\lim _{b\to 0 }=0.5$ to $\lim _{b\to \infty }H(b)=0$ so that the population approaches maximum relative poverty as the shape parameter gets smaller, and approaches zero relative poverty as it gets larger.  For the latter, the Weibull collapses to a single point, which is reflective of all incomes being equal.

\subsubsection{Exponential}The Exponential model appears in the last family with $b =1$ and the  previous Pareto family as a limiting case
   when $a\to \infty .$  It has $H=1-\sqrt 2\,/2.$

\subsection{Example of transference of income to reduce $H$ to 0.}

Suppose the government of Hong Kong wants to bring the income of all those below the relative poverty line $L$
up to $L$; this requires an amount $T=L-\e [X|X<L].$  It proposes to do this by imposing a flat rate of $r$\% on those with income above some $c>M$ and transfers the total proceeds to pay for $T$. What choice of $c$ and $r$ will achieve this aim?

Formally, the following transfer function $Y=t(X)$ is proposed:
\begin{equation}\label{transfer}
    t(x)= \left\{
            \begin{array}{ll}
           L  , & \hbox{if $x\leq L$\; ;} \\
              x, & \hbox{if $L<x\leq c$\; ;} \\
             (1-r)x, & \hbox{if $c <x $\;.}
            \end{array}
          \right.
\end{equation}
The effect on $H$ of such a transfer is changing $H_X=0.244$ to $H_Y=0.$

As an example, assume $F$ is standard lognormal and so $\e [X|X<L]= 0.3053757,$ found by numerical integration.
Therefore a total amount of $T= L-\e [X|X<L]=0.1946243$ must be found from those having incomes above $c$.
The 20\% of the largest incomes are above $c=x_{0.8}=2.32$. Therefore we must choose $r$ so that $0.1946243=T=r\,\e [X|X>2.32]= 4.635984\, r,$ so $r= 0.042.$   Different combinations of $c$, $r$ could be found which achieve the same
result.    Then those with incomes greater than 2.32 times
the median $M=1$ are taxed at the rate of 4.2\%, and  this will reduce to 0 the proportion below the relative poverty line.  Of course this is a theoretical example in that we are assuming a lognormal model for Hong Kong incomes, and do 
not include the costs of implementing such a tax. Nevertheless, it gives an indication of what can be done.
It should be possible to implement a distribution-free version of this transference example based on a large sample of data that are well-modeled by $F$.

\section{Inference for $H_p$}\label{sec:inference}

In the section we consider estimation of $H_p$ when the complete data set is available, but also for data available in grouped format only.  The latter is important since income and similar data are often not available in full detail to protect confidentiality.

\subsection{Point estimators using the complete data set}

Let $\widehat{M}$ denote the sample median estimator for a random sample of size $n$ denoted $X_1,\ldots,X_n$. For $I(\cdot)$ denoting the indicator function which is equal to one if its argument is true and false otherwise, a simple estimator of $H_p$ is
\begin{equation}
    \widehat{H}_p= \frac{1}{n}\sum^n_{i=1}I(X_i \leq p\times \widehat{M}).
    \label{Hhat}
\end{equation}

While one may mistake $\widehat{H}_p$ as a simple estimator of a population proportion, a complicating factor is that $\widehat{M}$, itself an estimator, is also random. Inferential properties of $\widehat{H}_p$, including the mean and variance, are therefore not straightforward.  Therefore we find the approximate mean and variance of the estimator, which depend on the variance of the sample median, to see whether these are useful in inference.  Therefore we find the approximate mean and variance of $$\tilde H_p=F(p\widehat{M})$$ to see whether these are useful in inference.  First, we note that, $\text{Var}(\widehat{M})\doteq 1/[4f^2(M)]$ \citep[e.g.][]{Das-2006}.

 The mean of the estimator $\tilde H$ is approximately given in terms of $L=pM$, $F$ and its assumed density $f=F'$ and second derivative  $f'=F''$ at $L$ by
 \begin{equation}\label{meanHtilde}
   \e [\tilde H_p] = \e [F(p\hat M)]\doteq  H_p +\frac {f'(L)}{2} \var (p\widehat M])
   \doteq H_p+  \frac{p^2}{8n}\;\frac {f'(L)}{f^2(M)}~.
  \end{equation}
 This shows that the bias of $\widehat H_p$ is of order $1/n$ for large $n$.

 The asymptotic variance (ASV) of $\tilde H_p$ is determined by:
 \begin{equation}\label{varHtilde}
n\var (\tilde H_p) = n\,\var [F(p\widehat M)]\doteq  n\,f^2(L)\var (p\widehat M) \doteq \frac {p^2f^2(L)}{4 f^2(M)}
 \equiv \sigma ^2~.
  \end{equation}
Consequently, the variance of $\widehat{H}_p$ can be computed using \eqref{varHtilde} and estimates of $M$ and $f$. However, our simulations, some of which are reported later, revealed that when constructing confidence intervals, this approximate asymptotic variance is not a good choice for some distributions.  Therefore, a better choice may be to use the formulation provided by \cite{zheng2001statistical} which is, using our notation,
\begin{equation}\label{ASV}
   n\,\var (\widehat{H}_p) \doteq H_p(1-H_p) - pH_p \frac{f(L)}{f(M)} + \frac {p^2f^2(L)}{4 f^2(M)}=H_p(1-H_p)-2H_p\sigma + \sigma^2.
\end{equation}
The difference between approximate ASV in \eqref{varHtilde} and the ASV in \eqref{ASV} is the first two terms in \eqref{ASV} given as $H(1-H)-2H\sigma$.  For many distributions this is negative and therefore decreases the ASV resulting in less conservative confidence intervals.

\subsection{Point estimators from grouped data}\label{sec:grouped}

To protect privacy and confidentiality, income data is often only available in summary, or grouped format either presenting values for income quantiles (e.g. deciles) or frequency of incomes within bins.  When the mean income is available within bins, \cite{lyon2016advantages} provides estimators of $G$ using linear interpolation within bins and an exponential tail.  When bin means are not available, another possibility is to approximate the income distribution using percentile matching methods such as those available in the \text{bda} package \citep{wang2015bda} in R \citep{R}.  \cite{Dedduwakumara&Prendergast18,dedduwakumara2019interval} used the \cite{lyon2016advantages} linear interpolation method and percentile matching for the  FKML parameterization of the Generalized Lambda Distribution \citep{freimer1988study} to obtain interval estimators for quantiles and inequality measures  respectively.  With four parameters, including two shape parameters, the GLD is capable of approximating many other distributions, including those that are often used to model income data.  Consequently, the GLD is a popular choice in financial modeling \citep[see e.g.,][]{pfaff2016financial}.  We therefore use these approaches to obtain estimates of $H$ when grouped data are available.

In what follows, let  $\widehat{F}_g$ and $\widehat{f}_g$ denote estimators to $F$ and $f=F'$ (the density function) using only data available in grouped format.  As noted above, we will use two different approaches although other estimators may be similarly applied.  Our estimate for $H_p$ from our grouped data density estimate is then
\begin{equation}
    \widehat{H}_{g,p}=\widehat{F}_g\left(\widehat{m}_g\times p\right)\label{Hp}
\end{equation}
where $\widehat{m}_g=\widehat{F}_g^{-1}(0.5)$ is the estimated median.

\subsubsection{The linear interpolation method}

Hereafter we adopt the notation of \cite{lyon2016advantages} to describe a grouped data set except
where it differs with ours already introduced above.  Given $J$ ordered intervals (or bins), denoted $[a_{j-1},a_j)$ with midpoint $x_j^c$, $j=1,\ldots,J$,  the proportion of values falling within the $j$th interval is called 
$q_j$. We further assume that the mean of the data falling within the $j$th interval is available and denoted by $\overline{x}_j$.  By using both the midpoint and mean, an estimate to the density within the $j$th interval can be approximated by the linear equation
\begin{equation*}
    \widehat{f}_j(x)=\widehat{\alpha}_j+\widehat{\beta}_j x
\end{equation*}
for $x\in[a_{j-1},a_j)$ and
\begin{equation*}
\widehat{\beta_j}=q_j\frac{12(\overline{x}_j-x^c_j)}{(a_j-a_{j-1})^3},\
\widehat{\alpha}_j=\frac{q_j}{a_j-a_{j-1}}-\widehat{\beta_j}x^c_j.
\end{equation*}

It is natural to consider the final $J$th interval as being unbounded since the maximum value in the data set is not likely to be the largest value in the population.  \cite{lyon2016advantages} suggest using an exponential tail estimate given as
\begin{equation}
\widehat{f}_J(x)=\frac{q_J}{\overline{x}_J-a_{J-1}}\exp\left\{-\frac{(x-a_{J-1})}{(\overline{x}_J-a_{J-1})}\right\}
\label{fJ}
\end{equation}
and this choice appears to work well.

Using the linear equations above, and exponential tail for the final interval, the linear interpolation estimate for $f$ is then
\begin{equation}
    \widehat{f}_g(x)=\begin{cases}
    \widehat{\alpha}_1+\widehat{\beta}_1 x,& x\in[a_0,a_1)\\
    \widehat{\alpha}_2+\widehat{\beta}_2 x,& x\in[a_1,a_2)\\
    \vdots & \vdots\\
    \widehat{f}_J(x),& x\in[a_{J-1},\infty)
    \end{cases}
\end{equation}
where $\widehat{f}_J(x)$ is given in \eqref{fJ}.

\subsubsection{The GLD percentile matching method}\label{sec:GLDmethod}

The FKML parameterization \citep{freimer1988study} of the GLD is perhaps the most useful since, unlike others, it is defined for all paramater choices with the exception of requiring a positive scale.  For $\lambda$ and $\eta$ denoting location and inverse-scale parameters, and $\alpha$ and $\beta$ shape parameters, the GLD quantile function is given as
\begin{equation}
Q(u)=\lambda + \frac{1}{\eta} \bigg[\frac{(u^\alpha-1)}{\alpha}-\frac{(1-u)^\beta-1}{\beta}\bigg].
\label{GLD}
\end{equation}

With quantiles available in grouped data (e.g. as deciles or as bounds of bins), there exist percentile matching methods to estimate the GLD parameters \citep[see][]{karian1999fitting,tarsitano2005estimation} and we use the functionality for this provided in the \texttt{bda} package \cite{wang2015bda}.  We then use those estimates the GLD quantile and distribution functions provided in the \texttt{gld} R package \citep{gld} as our estimated functions.

\subsection{Confidence interval estimators }\label{sec:theoryHohat}

In this section we provide several approximate confidence intervals for $H_p$ for a given $p$.  These are approximate intervals due to the difficulty in obtaining the exact variance for the $H_p$ estimator.
Throughout let $z_{1-\alpha/2}$ denote the $1-\alpha/2$ percentile from the standard normal distribution.  We begin with the simplest possibility of using the usual asymptotic interval for a binomial proportion.

\subsubsection{The asymptotic Wald-type interval for a binomial proportion}\label{sec:bino}

One possibility is to assume that the median is in fact known, and to then use a confidence interval for a binomial proportion. For example, a simple candidate is an approximate $(1-\alpha/2)\times 100$ confidence interval for $H_p$ as $\widehat{H}_p \pm z_{1-\alpha/2} \sqrt{\widehat{H}_p(1-\widehat{H}_p)/n}$. 

Further, there are numerous alternative confidence intervals which are proposed in the literature for binomial proportions. Several such alternative confidence intervals are the Agresti-Coull interval \citep{agresti1998approximate}, the Pearson-Clopper interval\citep{clopper1934use} and the Wilson interval \citep{wilson1927probable}. An extensive overview for some these methods are provided in \cite{brown2001interval}.

\subsubsection{Wald-type intervals with approximate standard error}\label{sec:wald}

Let $q(u)=1/f(Q(u))$ for $u\in [0,1]$.  Then $q(u)$ is called the quantile density function \citep{par-1979}
and estimators for it in the form of a kernel density estimator have been 
studied by \cite{welsh-1988,jones-1992,prendergast2016exploiting}. They suggest a bandwidth $b$ for the estimator that minimizes the asymptotic mean square error. We denote this estimator by 
$\widehat{q}(u)$.   This estimator can be written
\begin{equation}\label{qu}
    \widehat{q}(u)=\sum^n_{i=1}X_{[i]}\left[k_b(u-(i-1)/n)-k_b(u-i/n)\right]
\end{equation}
where $k_b$ is the \cite{epan-1969} kernel with bandwidth $b$ and $X_{[1]}\leq X_{[1]}\leq \ldots\leq X_{[n]}$ are the ordered $X_i$s.

Note that, from \eqref{varHtilde} and \eqref{ASV}, choices for an approximate standard error for $\widehat{H}_p$ are
\begin{equation*}
    \text{SE}_1\approx \frac{p\widehat{q}(0.5)\widehat{f}(p\widehat{M})}{2\sqrt{n}},\;\; \text{SE}_2=\sqrt{\text{SE}^2_1 + \frac{\widehat{H}_p(1-\widehat H)_p}{n}-2\widehat{H}_p\frac{\text{SE}_1}{\sqrt{n}}}\label{SE}
\end{equation*}
where $\widehat{f}$ is an estimate of the density function, $f$ and $\widehat{q}(0.5)$ is estimated using \eqref{qu}.  To estimate $f$, we use standard R \texttt{density} function which is also a kernel density estimator with a Gaussian kernel.  Then, an approximate $(1-\alpha/2)\times 100$ confidence interval is $$\widehat{H}_p \pm z_{1-\alpha/2}\times \text{SE}$$
where SE is either SE$_1$ or SE$_2$ above.

\subsubsection{Substitution intervals using a confidence interval for the median estimator}\label{sec:sub}

Let $[M_l,\ M_u]$ denote a confidence interval for the median.  Then another possibility is to construct a confidence interval for $H$ by $[\widehat{F}(M_l/2),\ \widehat{F}(M_u/2)$.  As noted in Section \ref{sec:inference}, the variance of the median estimator is approximately $1/[4f^2(M)]$ and we can estimate the $1/f(M)$ by $\widehat{q}(0.5)$ from \eqref{qu}.  Therefore, an approximate $(1-\alpha/2)\times 100$\% confidence for $M$ that may be used is $\widehat{M}\pm z_{1-\alpha/2}\widehat{q}(0.5)/(2\sqrt{n})$.  Other confidence intervals for the median could also be used.  As our estimate to $F$, we use the empirical cumulative distribution function.

\subsubsection{Bootstrap intervals}\label{sec:boot}

For comparison with the Wald-type intervals, we also consider bootstrap confidence intervals.  Due to superior computational efficiency, for our simulations we present the results from percentile bootstrapping for which 500 samples of size $n$ are drawn from the data set, also of size $n$.  Then, in our simulations 95\%, confidence intervals are the 2.5\% and 95.5\% percentiles from the 500 estimates of $H$ obtained from the samples.  As we shall see, very good coverages are achieved for this approach although in practice one may wish to employ what many consider to be superior bootstrapping methods at the expense of efficiency. An example would be the BCa method \citep{efron1987better}. For grouped data, we conduct the bootstrapping by sampling as above but from the estimated quantile functions arising from the density estimates from Section \ref{sec:grouped}.

\subsection{Simulation studies of the  estimator of $H$}\label{sec:simuHhat}

We start by presenting the coverage probabilities for the Binomial proportion confidence intervals, Wald-type confidence intervals and percentile bootstrap confidence intervals. For the Singh-Maddala distribution we consider the  parameter values $a=1.6971$, $b=87.6981$ and $q=8.3679$ reported by \cite{mcdonald1984some} which was fitted to a data set of US family incomes. The Dagum distribution is also considered and with the parameter choices of $a=4.273$ $b=14.28$ and $p=0.36$ which were used in \cite{kleiber2008guide}, used to model US family incomes sampled in 1969.  The Dagum and Singh-Maddala distributions are commonly referred to as income distributions, hence our inclusion here, and more on how they are related can be be found in \cite{kleiber1996dagum}.  We also focus our attention on $p=0.5$ and $p=0.6$ although similar results were also achieved for $p=0.4$.

\begin{table}[h!t]
\footnotesize
\renewcommand{\tabcolsep}{1.5mm}
  \centering
  
    \caption{Empirical coverage probabilities and average widths (in brackets) for Wald-type asymptotic confidence intervals and percentile bootstrap confidence interval estimates of $H_p$ with $p=0.5$ and $p=0.6$, estimated at nominal level 95\%, each based on 1000 replications and 500 bootstrap replicates.\strut}
\label{tab3}
\hspace*{-1cm}
\begin{tabular}{ccccccc}
\toprule
Method & \textit{p} & \textit{F} & \textit{n=100} & \textit{n=250} & \textit{n=500} & \textit{n=1000}\\
\midrule

&& LN$(0,1)$ & 0.963 (0.167) & 0.975 (0.106) & 0.973 (0.075) & 0.969 (0.053)\\
&& EXP$(1)$ & 0.977 (0.177) & 0.986 (0.113) & 0.986 (0.080) & 0.990 (0.056)\\
&0.5& Pareto$(1)$ & 0.994 (0.184) & 0.994 (0.117) & 0.994 (0.083) & 0.989 (0.058)\\
&& Pareto$(2)$ & 0.977 (0.181) & 0.995 (0.115) & 0.987 (0.081) & 0.991 (0.057)\\
&& Dagum & 0.969 (0.148) & 0.970 (0.095) & 0.976 (0.067) & 0.976 (0.048)\\
Binomial&& Singh-Maddala & 0.957 (0.154) & 0.970 (0.099) & 0.972 (0.070) & 0.971 (0.049)\\
\cmidrule{2-7}
&& LN$(0,1)$ & 0.977 (0.179) & 0.987 (0.114) & 0.978 (0.081) & 0.978 (0.057)\\
&& EXP$(1)$ & 0.993 (0.185) & 0.991 (0.117) & 0.992 (0.083) & 0.996 (0.059)\\
&0.6& Pareto$(1)$ & 1.000 (0.189) & 0.999 (0.120) & 0.999 (0.085) & 0.998 (0.060)\\
&& Pareto$(2)$ & 0.991 (0.187) & 0.992 (0.119) & 0.995 (0.084) & 0.996 (0.059)\\
&& Dagum & 0.969 (0.166) & 0.973 (0.105) & 0.977 (0.075) & 0.983 (0.053)\\
&& Singh-Maddala & 0.974 (0.170) & 0.981 (0.108) & 0.988 (0.077) & 0.985 (0.054)\\
\midrule

&& LN$(0,1)$ & 0.962 (0.158) & 0.950 (0.097) & 0.947 (0.068) & 0.951 (0.048) \\
&& EXP$(1)$ & 0.969 (0.161) & 0.974 (0.098) & 0.977 (0.068) & 0.966 (0.047) \\
&0.5& Pareto$(1)$ & 0.976 (0.171) & 0.973 (0.100) & 0.968 (0.070) & 0.969 (0.055)\\
&& Pareto$(2)$ & 0.984 (0.173) & 0.981 (0.105) & 0.980 (0.073) & 0.980 (0.052) \\
&& Dagum & 0.884 (0.106) & 0.837 (0.062) & 0.814 (0.042) & 0.828 (0.029) \\
Wald(1)&& Singh-Maddala & 0.923 (0.124) & 0.891 (0.073) & 0.877 (0.05) & 0.869 (0.034)\\
\cmidrule{2-7}
&& LN$(0,1)$ & 0.974 (0.163) & 0.959 (0.098) & 0.955 (0.068) & 0.956 (0.048) \\
&& EXP$(1)$ & 0.979 (0.161) & 0.977 (0.096) & 0.968 (0.066) & 0.966 (0.045) \\
&0.6& Pareto$(1)$ & 0.986 (0.171) & 0.976 (0.099) & 0.970 (0.071) & 0.982 (0.055)\\
&& Pareto$(2)$ & 0.990 (0.172) & 0.990 (0.104) & 0.982 (0.071) & 0.991 (0.050) \\
&& Dagum & 0.879 (0.115) & 0.858 (0.067) & 0.838 (0.045) & 0.817 (0.031)\\
&& Singh-Maddala & 0.941 (0.131) & 0.901 (0.077) & 0.884 (0.053) & 0.888 (0.036) \\
\midrule

&& LN$(0,1)$& 0.926 (0.140) & 0.936 (0.090) & 0.920 (0.064) & 0.936 (0.046)\\
&& EXP$(1)$&  0.938 (0.137) & 0.966 (0.087) & 0.930 (0.062) & 0.952 (0.044)\\
&0.5& Pareto$(1)$ & 0.944 (0.131) & 0.938 (0.084) & 0.954 (0.061) & 0.936 (0.045)\\
&& Pareto$(2)$ & 0.960 (0.134) & 0.930 (0.085) & 0.968 (0.061) & 0.926 (0.043)\\
&& Dagum & 0.936 (0.133) & 0.934 (0.085) & 0.950 (0.060) & 0.946 (0.042)\\
Wald(2)&& Singh-Maddala & 0.920 (0.136) & 0.934 (0.086) & 0.942 (0.061) & 0.946 (0.043)\\
\cmidrule{2-7}
&& LN$(0,1)$& 0.926 (0.138) & 0.946 (0.089) & 0.954 (0.063) & 0.948 (0.045)\\
&& EXP$(1)$ & 0.962 (0.131) & 0.940 (0.083) & 0.950 (0.059) & 0.962 (0.042)\\
&0.6& Pareto$(1)$ & 0.934 (0.121) & 0.950 (0.077) & 0.946 (0.057) & 0.940 (0.043)\\
&& Pareto$(2)$ & 0.962 (0.127) & 0.924 (0.081) & 0.944 (0.057) & 0.946 (0.041)\\
&& Dagum & 0.950 (0.139) & 0.932 (0.088) & 0.962 (0.062) & 0.944 (0.044)\\
&& Singh-Maddala & 0.942 (0.140) & 0.966 (0.089) & 0.958 (0.063) & 0.954 (0.045)\\

\midrule

&& LN$(0,1)$  & 0.984 (0.162) & 0.975 (0.101) & 0.978 (0.071) & 0.966 (0.050)\\
&& EXP$(1)$ & 0.982 (0.152) & 0.972 (0.094) & 0.976 (0.066) & 0.966 (0.046)\\
&0.5& Pareto$(1)$ & 0.993 (0.154) & 0.987 (0.095) & 0.98 (0.066) & 0.963 (0.046)\\
&& Pareto$(2)$ & 0.990 (0.154) & 0.974 (0.095) & 0.967 (0.066) & 0.975 (0.046)\\
&& Dagum & 0.968 (0.140) & 0.972 (0.088) & 0.963 (0.062) & 0.960 (0.044)\\
Bootstrap&& Singh-Maddala & 0.983 (0.146) & 0.966 (0.092) & 0.966 (0.064) & 0.967 (0.045)\\
\cmidrule{2-7}
&& LN$(0,1)$ & 0.987 (0.161) & 0.975 (0.099) & 0.979 (0.069) & 0.973 (0.048)\\
&& EXP$(1)$  & 0.991 (0.149) & 0.983 (0.091) & 0.975 (0.063) & 0.970 (0.044)\\
&0.6& Pareto$(1)$ & 0.998 (0.144) & 0.984 (0.089) & 0.986 (0.061) & 0.978 (0.043)\\
&& Pareto$(2)$ & 0.996 (0.147) & 0.988 (0.090) & 0.987 (0.062) & 0.971 (0.043)\\
&& Dagum & 0.975 (0.149) & 0.970 (0.092) & 0.964 (0.065) & 0.963 (0.046)\\
&& Singh-Maddala & 0.984 (0.152) & 0.965 (0.095) & 0.967 (0.066) & 0.953 (0.046)\\

\bottomrule
\end{tabular}
 \end{table}

From Table \ref{tab3}, for all the distributions and choices of $p$, the binomial proportion intervals from Section \ref{sec:bino} produce conservative confidence intervals with coverage exceeding the nominal level of 0.95, but with mean widths that suggest the intervals may still be useful in practice. The Wald-type interval using SE$_1$ from Section \ref{sec:wald}, labeled Wald(1) in the table, generally provide good coverage, tending toward conservative and slightly narrower than the intervals for the binomial intervals.  However, the coverages are too low for the Dagum and Singh-Maddala distributions, even for $n=1000$.  However, coverages are typically very good when using SE$_2$, labeled Wald(2), based on the asymptotic variance results from \cite{zheng2001statistical} with all coverages between 0.92 and 0.97, and with most very close to the nominal 0.95. Widths were usually narrower compared to when using SE$_1$, except for the Dagum and Sing-Maddala intervals which were too liberal when using SE$_1$.   The percentile bootstrap confidence intervals were mildly conservative, but with coverages not too far from nominal.   

For simplicity we present the coverages for other methods in Appendix \ref{Other CI's}, Tables \ref{tab:other_p5} and \ref{tab:other_p6}, and we briefly summarise the results here.  The coverages for the alternative binomial proportion interval estimators mentioned in Section \ref{sec:bino} are similarly as conservative than those for the standard interval, and with similar widths.  Performance is mixed for the interval estimator whereby the interval for the median is substituted into the empirical distribution function.  Poor coverages are observed for the Dagum and Singh-Maddala distributions for $p=0.5$ and for the Dagum when $p=0.6$.  However, the other coverages are good and typically less conservative than the binomial proportion estimators.  .

Overall, the best performing interval was the Wald interval using SE$_2$ which while not always conservative, produced coverages close to nominal.  The Wald interval using SE$_2$ was generally good but with coverages more conservation for some distributions, and too low for others.  The bootstrap intervals provided conservative results but with narrower intervals than the simple binomial method.

\begin{table}[h!t]                                             
\footnotesize
\renewcommand{\tabcolsep}{1.5mm}
  \centering
  
    \caption{This table shows a comparison of the mean standard error and the absolute bias of the estimates of $H_p$ from data grouped summarized using deciles, and when using the fitted GLD and linear interpolation (LI). These mean values were calculated for 500 fitting results. \strut}
\label{tab4}
\hspace*{-1cm}
\begin{tabular}{ccccccc}
\toprule
Method &\textit{p} & \textit{F} & \textit{n=100} & \textit{n=250} & \textit{n=500} & \textit{n=1000}\\
\midrule
&& LN$0,1$ & 0.001 (0.001) & 0.000 (0.006) & 0.000 (0.009) & 0.000 (0.010)\\
&& EXP$(1)$ & 0.001 (0.009) & 0.000 (0.001) & 0.000 (0.002) & 0.000 (0.003)\\
&0.5&  Pareto$(1)$ & 0.004 (0.016) & 0.002 (0.012) & 0.003 (0.010) & 0.002 (0.008)\\
&& Pareto$(2)$ & 0.001 (0.001) & 0.000 (0.002) & 0.000 (0.003) & 0.000 (0.005)\\
&& Dagum & 0.001 (0.014) & 0.000 (0.007) & 0.000 (0.004) & 0.000 (0.003)\\
GLD&& Singh-Maddala & 0.001 (0.010) & 0.000 (0.001) & 0.000 (0.002) & 0.000 (0.003)\\
\cmidrule{2-7}
&& LN$0,1$ & 0.001 (0.001) & 0.000 (0.005) & 0.000 (0.008) & 0.000 (0.009)\\
&& EXP$(1)$ & 0.000 (0.009) & 0.000 (0.002) & 0.000 (0.001) & 0.000 (0.003)\\
&0.6&  Pareto$(1)$ & 0.003 (0.019) & 0.003 (0.013) & 0.002 (0.010) & 0.003 (0.011)\\
&& Pareto$(2)$ & 0.001 (0.004) & 0.000 (0.001) & 0.000 (0.003) & 0.000 (0.004)\\
&& Dagum & 0.001 (0.013) & 0.000 (0.005) & 0.000 (0.002) & 0.000 (0.003)\\
&& Singh-Maddala & 0.001 (0.008) & 0.000 (0.000) & 0.000 (0.002) & 0.000 (0.004)\\
\hline
&& LN$0,1$ & 0.002 (0.002) & 0.001 (0.000) & 0.000 (0.001) & 0.000 (0.000)\\
&& EXP$(1)$ & 0.001 (0.002) & 0.000 (0.001) & 0.000 (0.000) & 0.000 (0.000)\\
&0.5&  Pareto$(1)$ & 0.001 (0.004) & 0.000 (0.002) & 0.000 (0.002) & 0.000 (0.000)\\
&& Pareto$(2)$ & 0.001 (0.001) & 0.000 (0.002) & 0.000 (0.001) & 0.000 (0.001)\\
&&Dagum & 0.001 (0.000) & 0.000 (0.000) & 0.000 (0.000) & 0.000 (0.000)\\
LI && Singh-Maddala & 0.001 (0.001) & 0.000 (0.002) & 0.000 (0.000) & 0.000 (0.001)\\
\cmidrule{2-7}
&& LN$0,1$ & 0.001 (0.004) & 0.000 (0.001) & 0.000 (0.001) & 0.000 (0.001)\\
&& EXP$(1)$ & 0.001 (0.002) & 0.000 (0.001) & 0.000 (0.001) & 0.000 (0.000)\\
&0.6&  Pareto$(1)$ & 0.001 (0.001) & 0.000 (0.001) & 0.000 (0.001) & 0.000 (0.001)\\
&& Pareto$(2)$ & 0.001 (0.002) & 0.000 (0.001) & 0.000 (0.000) & 0.000 (0.000)\\
&&Dagum & 0.001 (0.002) & 0.000 (0.002) & 0.000 (0.000) & 0.000 (0.000)\\
&&Singh-Maddala  & 0.001 (0.002) & 0.001 (0.000) & 0.000 (0.001) & 0.000 (0.001)\\

\bottomrule
\end{tabular}
 \end{table}

In Table \ref{tab4} we consider estimation of $H_p$ from grouped data using the density estimates provided in Section \ref{sec:grouped} and when data is summarized using deciles. A low  mean squared error and small absolute bias for all the sample sizes and distributions found.  Results were improved using the linear interpolotion approach although an advantage of the GLD method is that group means are not needed.  Hence, in practice if only counts are provided, then the GLD density approximation can be used to obtain good results, although if means are also available then linear interpolation can be used to improve the estimates, 

We also assessed the performance of confidence intervals for $H_p$ from grouped data using bootstrapping where the samples were generate from the estimated income distribution (see Section \ref{sec:grouped}).  For simplicity we provide the results as Table \ref{tab:bootCI_grouped} in the Appendix. For the GLD approach, coverages were conservative to good for most distributions, with improvement for larger sample sizes.  However, coverage was low for the Pareto(1) distribution, even for $n=1000$.  Conversely, coverages are very good for the linear interpolation approach, including for the Pareto(1) distribution.  The Pareto densities have a steep downward gradient at $L=M/2$, and the improvement for the linear interpolation approach suggests that additional local information around this point, in this case the bin mean, can greatly improve performance.  This improvement was seen also for the Pareto(2) distribution, but in a different way, where the GLD approach was instead very conservative.  Overall, however, bootstrapping with an estimated density from grouped data appears to be a good option, with preference towards linear interpolation if bin means are available. 

\section{Applications}\label{sec:Applications}

In this section we consider application of the measure to two data sets.

\subsection{Application 1: Earnings data } 

The data we consider here includes the hourly earnings of males and females in the US in 1992 (2962 observations; 1371 females and 1591 males) and 1998 (2603 observations; 1210 females
and 1393 males) and is available in the \textit{Ecdat} package \citep{croissant2016Ecdat}. This has also been previously considered in \cite{PR&ST18_QRI} to compare income inequality measures.  

\begin{table}[h!t]
\small
  \centering
    \caption{Point and interval estimates of $H_{0.5}$ for earnings of males (M) and females (F) in 1992 and 1998 including differences between years (labeled 1998-1992) and between gender (labeled M-F). CI-W refers to Wald(2) intervals and CI-B to bootstrap intervals.
}
\label{tab5}
\hspace*{-1cm}
\begin{tabular}{cccc}
\toprule
Year &Gender && \textit{$H_{0.5}$} \\
\midrule
1992&M & Est. & 0.083 \\
&& CI-W&  (0.070, 0.096) \\
&& CI-B &  (0.071, 0.100) \\
&F&Est.& 0.053 \\
&&CI-W & (0.041, 0.064) \\
&&CI-B & (0.040, 0.065) \\
&M-F & Est. & 0.030 \\
&&CI-W & (0.013, 0.048) \\
&&CI-B & (0.014, 0.053) \\

\addlinespace

1998&M & Est. & 0.072 \\
&& CI-W & (0.058, 0.085) \\
&& CI-B & (0.056, 0.086) \\
&F&Est. & 0.072 \\
&&CI-W & (0.058, 0.086) \\
&&CI-B & (0.056, 0.085) \\
&M-F &Est.& 0 \\
&&CI-W& (-0.020, 0.019) \\
&&CI-B& (-0.022, 0.018) \\

\addlinespace

1998-1992&M & Est. & -0.011\\
&& CI-W & (-0.030, 0.008)\\
&& CI-B & (-0.034, 0.006)\\
&F&Est. & 0.019\\
&&CI-W & (0.001, 0.038)\\
&&CI-B & (-0.003, 0.037)\\
&M-F &Est. & -0.031\\
&&CI-W & (-0.057, -0.004)\\
&&CI-B & -\\

\bottomrule
\end{tabular}
 \end{table}

In Table \ref{tab5} we present the point and interval estimates for head count ratio, $H_{0.5}$, for both males and females in 1992 and 1998 and also for the difference between genders and difference between years (1998-1992). Both the bootstrap and Wald-type intervals produce similar results for the $H_{0.5}$ with narrow interval widths. Further it suggests there is a difference in poverty between males and females for the year 1992. For 1998, $H_{0.5}$ suggests little difference between genders. There is no significant difference between males from 1992-1998 as suggested by $H_{0.5}$  but for females it is significant according to the measure.

\subsection{Application 2: Australian disposable weekly income} 

\begin{table}[h!t]
\small
  \centering
    \caption{Equivalized disposable weekly income (DWI) in Australian dollars adjusted for inflation to 2013-2014 dollars, for selected financial years. The entries represent thousands of people.}
\label{tab6}
\hspace*{-1cm}
\begin{tabular}{cccccc}
\toprule
 & 2003-2004 & 2005-2006 & 2008-2010 & 2011-2012 & 2013-2014\\
\midrule
\lbrack0,0\rbrack & 87.3 & 73.7 & 89.0 & 87.4 & 86.4\\
\lbrack1,49\rbrack & 94.1 & 90.1 & 95.8 & 81.6 & 95.3\\
\lbrack50,99\rbrack & 49.7 & 63.1 & 61.3 & 85.3 & 78.9\\
\lbrack100,149\rbrack & 94.0 & 66.2 & 84.0 & 92.3 & 47.6\\
\lbrack150,199\rbrack & 129.9 & 108.6 & 125.1 & 107.3 & 134.9\\
\lbrack200,249\rbrack & 273.7 & 219.6 & 164.7 & 185.6 & 151.1\\
\lbrack250,299\rbrack & 657.6 & 443.7 & 351.5 & 335.0 & 373.4\\
\lbrack300,349\rbrack & 1385.5 & 1152.0 & 596.3 & 373.9 & 397.9\\
\lbrack350,399\rbrack & 1301.8 & 1187.5 & 1195.8 & 913.3 & 636.7\\
\lbrack400,449\rbrack & 1231.7 & 1111.8 & 1172.4 & 1184.1 & 1135.2\\
\lbrack450,499\rbrack & 1093.7 & 1052.3 & 933.4 & 1044.7 & 1175.2\\
\lbrack500,549\rbrack & 1043.0 & 1097.4 & 991.3 & 1019.7 & 1171.7\\
\lbrack550,599\rbrack & 1092.2 & 1057.0 & 1009.7 & 980.8 & 1093.0\\
\lbrack600,649\rbrack & 1087.5 & 1016.2 & 1046.4 & 926.3 & 956.6\\
\lbrack650,699\rbrack & 1083.5 & 1066.9 & 987.0 & 1021.9 & 972.7\\
\lbrack700,749\rbrack & 1092.8 & 1023.3 & 996.9 & 999.2 & 938.9\\
\lbrack750,799\rbrack & 959.9 & 834.1 & 1037.1 & 1038.1 & 1009.6\\
\lbrack800,849\rbrack & 878.1 & 940.4 & 829.3 & 989.4 & 1013.4\\
\lbrack850,899\rbrack & 718.3 & 828.5 & 806.5 & 959.7 & 1099.5\\
\lbrack900,949\rbrack & 612.2 & 746.6 & 793.0 & 896.4 & 826.2\\
\lbrack950,999\rbrack & 631.8 & 731.9 & 757.8 & 714.9 & 885.6\\
\lbrack1000,1049\rbrack & 506.8 & 547.5 & 630.3 & 690.1 & 692.6\\
\lbrack1050,1099\rbrack & 492.3 & 515.3 & 730.8 & 803.1 & 695.8\\
\lbrack1100,1199\rbrack & 750.3 & 933.9 & 1118.5 & 1245.7 & 1379.5\\
\lbrack1200,1299\rbrack & 529.4 & 674.2 & 906.1 & 985.3 & 1027.2\\
\lbrack1300,1499\rbrack & 706.4 & 863.9 & 1400.8 & 1499.2 & 1447.8\\
\lbrack1500,1699\rbrack & 387.9 & 469.6 & 889.7 & 995.4 & 938.5\\
\lbrack1700,1999\rbrack & 263.2 & 427.0 & 682.8 & 850.2 & 862.3\\
\lbrack2000,5000\rbrack & 371.9 & 588.4 & 1106.3 & 1082.9 & 1355.6\\
\hline
Total& 19,606.5 & 19,930.7 & 21,589.6 & 22,188.8 & 22,679.1\\
\bottomrule
\end{tabular}
\end{table}

In Table \ref{tab6} are the equalized disposal weekly incomes (DWI) in Australia reported by the Australian Bureau of Statistics \citep{australian2016household} for five years in grouped format. This data set has been previously looked at by \cite{PR&ST19_Decomp} to obtain estimates for a quantile-based measure of inequality. We consider two approaches to obtaining estimates of $H_p$ and their corresponding standard errors. In the first approach, we reconstruct the full data set from the available grouped data by simulating data values from the uniform distribution for the bounded intervals and for the final interval we generate data from the Pareto$(3)$ distribution. This method was used by \cite{PR&ST19_Decomp} to construct a complete data set for illustrative purposed. As the second approach, we use the GLD percentile matching method to estimate the underlying density of the grouped data as detailed in Section \ref{sec:GLDmethod}. 

\begin{table}[h!t]
\small
  \centering
    \caption{Estimates of the measure $H_{0.5}$ for the five income distributions from the constructed full data set and from the estimated GLD density. In parentheses are the values of bootstrap standard errors based on 500 re-samples.
}
\label{tab7}
\hspace*{-1cm}
\begin{tabular}{ccc}
\toprule
Year & Estimate using data & Estimate using GLD\\
\midrule
 2004 & 0.113 (0.001)& 0.110 (0.001)\\
2006 &0.116 (0.001)& 0.114 (0.001)\\
2010 &0.124 (0.001)& 0.121 (0.001)\\
2012 &0.119 (0.001)& 0.116 (0.001)\\
2014 &0.110 (0.001)& 0.109 (0.001)\\

\bottomrule
\end{tabular}
\end{table}

The estimates for each of the years are provided in Table \ref{tab7} based on both the constructed data and also from the fitted GLD. From these results, the estimated head count ratio, matched with small standard errors, suggests some degree of inequality over the years with a peak seen in 2010.        

\section{Discussion}\label{sec:discussion}

In this paper we have provide some insights and inference methods for the proportion of income earners less than a specific fraction of median income, commonly known as head count ratio. This measure is simple to understand and is widely used around the world.  We have provided examples for several probability distributions, including those that are often used to model income.

While the measure may be simple to understand, confidence intervals are not so straightforward due the need to both estimate the median and the income distribution function.  However, good coverages with narrow interval widths can be obtained using a Wald-type interval using a standard error approximated from the asymptotic results from \cite{zheng2001statistical}.  An alternative means to computing confidence intervals is to use the bootstrap approach.  Coverage for these intervals were typically close to nominal while being slightly conservative.  

Finally, we also showed that good estimates can be obtained from grouped data, which is commonly reported for incomes to protect privacy.  When group means are available, the linear interpolation approach by \cite{lyon2016advantages} is an attractive option.  When means are not available, we recommend approximating the income distribution using the GLD distribution and a percentile matching approach.  In both cases an estimate can then be obtained directly from the estimated quantile and distribution functions. 

\bibliography{relpov}
\bibliographystyle{authordate4}

\section{Appendix}\label{sec:appendix}






\subsection{Other Confidence Intervals}\label{Other CI's}

In these tables we provide coverages for other interval estimators of $H_p$.

\begin{table}[h]
\footnotesize
\renewcommand{\tabcolsep}{1.5mm}
  \centering
  
    \caption{Empirical coverage probabilities and average widths (in brackets) of confidence interval estimates of $H_p$ index with $p=0.5$, estimated at nominal level 95\% , each based on 1000 replications.  The first three are alternatives to the usual binomial proportion and the last is based on substitution of the CI for the median into the empirical distribution function.\strut}
\label{tab:other_p5}
\hspace*{-1cm}
\begin{tabular}{cccccc}
\toprule
\textit{CI} & \textit{F} & \textit{n=100} & \textit{n=250} & \textit{n=500} & \textit{n=1000}\\
\midrule

& LN$(0,1)$ & 0.972 (0.166) & 0.977 (0.106) & 0.969 (0.075) & 0.968 (0.053)\\
& EXP$(1)$ & 0.986 (0.175) & 0.982 (0.112) & 0.990 (0.079) & 0.991 (0.056)\\
Agresti-Coull & Pareto$(1)$ & 0.992 (0.181) & 0.994 (0.116) & 0.994 (0.082) & 0.989 (0.058)\\
& Pareto$(2)$ & 0.982 (0.179) & 0.994 (0.114) & 0.989 (0.081) & 0.993 (0.057)\\
& Dagum & 0.975 (0.150) & 0.972 (0.095) & 0.982 (0.067) & 0.977 (0.048)\\
& Singh-Maddala & 0.986 (0.154) & 0.983 (0.099) & 0.973 (0.070) & 0.975 (0.049)\\
\hline
& LN$(0,1)$ & 0.982 (0.175) & 0.982 (0.110) & 0.976 (0.077) & 0.973 (0.054)\\
& EXP$(1)$ & 0.986 (0.185) & 0.990 (0.116) & 0.991 (0.081) & 0.991 (0.057)\\
Pearson-Kloppe& Pareto$(1)$ & 0.998 (0.192) & 0.996 (0.120) & 0.994 (0.084) & 0.991 (0.059)\\
& Pareto$(2)$ & 0.993 (0.189) & 0.997 (0.118) & 0.989 (0.083) & 0.993 (0.058)\\
& Dagum & 0.984 (0.157) & 0.975 (0.099) & 0.985 (0.069) & 0.980 (0.049)\\
& Singh-Maddala & 0.989 (0.162) & 0.987 (0.102) & 0.973 (0.072) & 0.975 (0.050)\\
\hline
& LN$(0,1)$ & 0.972 (0.165) & 0.977 (0.106) & 0.969 (0.075) & 0.968 (0.053)\\
& EXP$(1)$ & 0.986 (0.174) & 0.982 (0.112) & 0.990 (0.079) & 0.991 (0.056)\\
Wilson& Pareto$(1)$ & 0.992 (0.181) & 0.994 (0.116) & 0.994 (0.082) & 0.989 (0.058)\\
& Pareto$(2)$ & 0.982 (0.178) & 0.994 (0.114) & 0.989 (0.081) & 0.993 (0.057)\\
& Dagum & 0.975 (0.148) & 0.972 (0.095) & 0.982 (0.067) & 0.977 (0.048)\\
& Singh-Maddala & 0.986 (0.153) & 0.977 (0.098) & 0.973 (0.070) & 0.975 (0.049)\\
\hline
& LN$(0,1)$ & 0.944 (0.157) & 0.958 (0.094) & 0.948 (0.070) & 0.946 (0.048)\\
& EXP$(1)$ & 0.962 (0.140) & 0.958 (0.085) & 0.934 (0.063) & 0.960 (0.044)\\
Median Est.& Pareto$(1)$ & 0.988 (0.178) & 0.976 (0.105) & 0.986 (0.080) & 0.982 (0.056)\\
& Pareto$(2)$ & 0.978 (0.158) & 0.972 (0.098) & 0.978 (0.070) & 0.968 (0.050)\\
& Dagum & 0.748 (0.084) & 0.804 (0.051) & 0.766 (0.037) & 0.804 (0.026)\\
& Singh-Maddala & 0.878 (0.105) & 0.814 (0.062) & 0.870 (0.047) & 0.850 (0.032)\\

\bottomrule
\end{tabular}
 \end{table}

\begin{table}[h]
\footnotesize
\renewcommand{\tabcolsep}{1.5mm}
  \centering
  
    \caption{Empirical coverage probabilities and average widths (in brackets) of confidence interval estimates of $H_p$ index with $p=0.6$, estimated at nominal level 95\% , each based on 1000 replications.  The first three are alternatives to the usual binomial proportion and the last is based on substitution of the CI for the median into the empirical distribution function.\strut}
\label{tab:other_p6}
\hspace*{-1cm}
\begin{tabular}{cccccc}
\toprule
\textit{CI} & \textit{F} & \textit{n=100} & \textit{n=250} & \textit{n=500} & \textit{n=1000}\\
\midrule

& LN$(0,1)$ & 0.988 (0.177) & 0.987 (0.113) & 0.984 (0.080) & 0.978 (0.057)\\
& EXP$(1)$ & 0.998 (0.182) & 0.992 (0.116) & 0.995 (0.083) & 0.998 (0.059)\\
& $\chi_1^2$ & 0.996 (0.188) & 1.000 (0.120) & 0.999 (0.085) & 0.998 (0.061)\\
Agresti-Coull& Pareto$(1)$ & 1.000 (0.186) & 1.000 (0.119) & 0.999 (0.085) & 0.998 (0.060)\\
& Pareto$(2)$ & 0.996 (0.184) & 0.996 (0.118) & 0.995 (0.084) & 0.998 (0.059)\\
& Dagum & 0.979 (0.165) & 0.976 (0.105) & 0.980 (0.075) & 0.984 (0.053)\\
& Singh-Maddala & 0.984 (0.169) & 0.984 (0.108) & 0.989 (0.076) & 0.986 (0.054)\\
\hline
& LN$(0,1)$ & 0.991 (0.187) & 0.989 (0.117) & 0.987 (0.082) & 0.979 (0.058)\\
& EXP$(1)$ & 0.998 (0.193) & 0.993 (0.121) & 0.995 (0.085) & 0.998 (0.060)\\
Pearson-Clopper& Pareto$(1)$ & 1.000 (0.197) & 1.000 (0.123) & 0.999 (0.087) & 0.998 (0.061)\\
& Pareto$(2)$ & 0.996 (0.195) & 0.996 (0.122) & 0.997 (0.086) & 0.998 (0.060)\\
& Dagum & 0.980 (0.174) & 0.976 (0.109) & 0.984 (0.077) & 0.985 (0.054)\\
& Singh-Maddala & 0.990 (0.178) & 0.984 (0.112) & 0.989 (0.078) & 0.986 (0.055)\\
\hline
& LN$(0,1)$ & 0.988 (0.176) & 0.987 (0.113) & 0.984 (0.080) & 0.978 (0.057)\\
& EXP$(1)$ & 0.998 (0.182) & 0.992 (0.116) & 0.995 (0.083) & 0.998 (0.059)\\
Wilson& Pareto$(1)$ & 1.000 (0.186) & 1.000 (0.119) & 0.999 (0.084) & 0.998 (0.060)\\
& Pareto$(2)$ & 0.996 (0.184) & 0.996 (0.118) & 0.995 (0.084) & 0.998 (0.059)\\
& Dagum & 0.979 (0.164) & 0.976 (0.105) & 0.980 (0.075) & 0.980 (0.053)\\
& Singh-Maddala & 0.984 (0.168) & 0.984 (0.108) & 0.989 (0.076) & 0.986 (0.054)\\
\hline
& LN$(0,1)$ & 0.970 (0.173) & 0.980 (0.104) & 0.980 (0.077) & 0.968 (0.054)\\
& EXP$(1)$ & 0.966 (0.158) & 0.972 (0.097) & 0.972 (0.068) & 0.988 (0.050)\\
Median Est.& Pareto$(1)$ & 0.998 (0.189) & 0.996 (0.111) & 0.994 (0.083) & 0.996 (0.058)\\
& Pareto$(2)$ & 0.994 (0.174) & 0.980 (0.105) & 0.992 (0.075) & 0.988 (0.054)\\
& Dagum & 0.876 (0.110) & 0.880 (0.067) & 0.858 (0.048) & 0.870 (0.034)\\
& Singh-Maddala & 0.934 (0.132) & 0.916 (0.078) & 0.938 (0.058) & 0.928 (0.040)\\

\bottomrule
\end{tabular}
 \end{table}

\subsection{Coverage probability with Grouped data}

In the following table are the coverage probabilities and average confidence widths for bootstrap confidence intervals for grouped data binned in to deciles with bin means. The underlying density of the grouped data are estimated using the linear interpolation method \citep{lyon2016advantages} and GLD method.  

\begin{table}[h]
\footnotesize
\renewcommand{\tabcolsep}{1.5mm}
  \centering
  
    \caption{Empirical coverage probabilities and average widths (in brackets) of bootstrap confidence interval estimates of $H_p$ ($p=0.5$) from data grouped in to deciles using the fitted density by GLD and Linear Interpolation method , estimated at nominal level 95\% , each based on 1000 replications and 500 bootstrap samples.}
\label{tab:bootCI_grouped}
\hspace*{-1cm}
\begin{tabular}{cccccc}
\toprule
Method & \textit{F} & \textit{n=100} & \textit{n=250} & \textit{n=500} & \textit{n=1000}\\
\midrule

&LN$(0,1)$ &0.988 (0.147)& 0.993 (0.094)& 0.989 (0.066)& 0.941 (0.047)\\
&EXP$(1)$ &0.981 (0.136)& 0.997 (0.086)& 0.997 (0.061)& 1 (0.043) \\
GLD&Pareto$(1)$&0.885 (0.133)& 0.853 (0.084)&  0.814 (0.060)& 0.768 (0.042) \\
&Pareto$(2)$ &0.966 (0.134)& 0.984 (0.085)&  0.983 (0.060)& 0.981 (0.043)\\
&Dagum &0.937 (0.131)& 0.975 (0.083)& 0.973 (0.059)& 0.976 (0.042) \\
&Singh-Maddala &0.974 (0.137)& 0.991 (0.087)& 0.993 (0.061) & 0.989 (0.044) \\
\addlinespace
&LN$(0,1)$ & 0.983 (0.155)& 0.978 (0.097)& 0.963 (0.069)& 0.959 (0.048)\\
&EXP$(1)$ & 0.978 (0.145)&  0.977 (0.090)& 0.971 (0.063)& 0.963 (0.044)\\
LI&Pareto$(1)$ & 0.988 (0.147)& 0.975 (0.091)& 0.972 (0.064)& 0.955 (0.045)\\
&Pareto$(2)$ & 0.986 (0.145)& 0.988 (0.091)& 0.965 (0.063)& 0.961 (0.044)\\
&Dagum & 0.945 (0.135)& 0.966 (0.085)&  0.963 (0.060)& 0.957 (0.042)\\
&Singh-Maddala & 0.971 (0.140)& 0.969 (0.088)& 0.968 (0.062)& 0.963 (0.043)\\

\bottomrule
\end{tabular}
 \end{table}

\end{document}